\documentclass[12pt]{article}
\usepackage{amssymb,amsmath,latexsym}
\usepackage{graphicx}
\usepackage[final]{pdfpages}
\usepackage{verbatim}

\usepackage{color}

\title{Gaussian fluctuations of products of random matrices \\ distributed close to the identity}

\author{Maxim Drabkin, Hermann Schulz-Baldes
\\
\\
{\small Department Mathematik,  Friedrich-Alexander Universit\"at Erlangen-N\"urnberg, }
\\
{\small Cauerstra{\ss}e 11, D-91058 Erlangen, Germany}
}

\date{ }

\newtheorem{theo}{Theorem}

\newtheorem{proposi}[theo]{Proposition}
\newtheorem{lemma}[theo]{Lemma}
\newtheorem{coro}[theo]{Corollary}

\newcommand{\CM}{{\mathbb C}}
\newcommand{\EM}{{\mathbb E}}
\newcommand{\NM}{{\mathbb N}}
\newcommand{\RM}{{\mathbb R}}
\newcommand{\SM}{{\mathbb S}}

\newcommand{\ZM}{{\mathbb Z}}

\newcommand{\Pp}{{\cal P}}
\newcommand{\PP}{{\bf P}}

\newcommand{\EE}{{\mathbb{E}}}

\newcommand{\Oo}{{\cal O}}
\newcommand{\Tr}{\mbox{\rm Tr}}

\newcommand{\Mm}{{\cal M}}

\newcommand{\Ll}{{\cal L}}
\newcommand{\Qq}{{\cal Q}}

\newcommand{\one}{{\bf 1}}

\newcommand{\Ran}{{\mbox{\rm Ran}}}

\begin{document}

\maketitle

\begin{abstract}
Products of random $2\times 2$ matrices exhibit Gaussian fluctuations around almost surely convergent Lyapunov exponents. In this paper, the distribution of the random matrices is supported by a small neighborhood of order $\lambda>0$ of the identity matrix. The Lyapunov exponent and the variance of the Gaussian fluctuations are calculated perturbatively in $\lambda$ and this requires a detailed analysis of the associated random dynamical system on the unit circle and its invariant measure. The result applies to anomalies and band edges of one-dimensional random Schr\"odinger operators.
\end{abstract}



\section{Main results}

The analysis of random finite difference equations such as harmonic chains or the one-dimensional Anderson model naturally leads to study products of random $2\times 2$ matrices. Asymptotically these products converge to a Gaussian process which is characterized by the Lyapunov exponent and its variance \cite{BL}. Often one is interested in a perturbative regime of small randomness. For example, the zero frequency limit of a harmonic chain is of this type, as well as the analysis of band center and band edges of the Anderson model \cite{Ish,KW,DG}. Also the behavior near the so-called critical energies of a random Kronig-Penney model is of this perturbative nature \cite{DKS}. It is the object of this work to develop a rigorously controlled pertubation theory for both the Lyapunov exponnet and its variance within a generic model covering all the situations alluded to above, and potentially others.

\vspace{.2cm}

More precisely, let us consider a family $(\mathcal{T}_{\lambda, n})_{n\geq 1}$ of random matrices in $\mbox{Sl}(2, \RM)$ which is of the form
\begin{equation}
\label{eq-matdef}
\mathcal{T}_{\lambda, n}
\;=\;
\exp\bigl(\lambda \mathcal{P}_n + \lambda^2 \mathcal{Q}_{\lambda,n}\bigr)
\;,
\end{equation}
where $\lambda\geq 0$ is a small coupling parameter and $\mathcal{P}_n,\,\mathcal{Q}_{\lambda,n}\in\mbox{sl}(2, \RM)$. The matrix $\mathcal{Q}_{\lambda,n}$ is supposed to be analytic in $\lambda$, and $\mathcal{P}_n$ and $\mathcal{Q}_{\lambda,n}$ are independent and identically distributed on a bounded set. The distribution of $\mathcal{T}_{\lambda, n}$ is thus supported in a neighborhood of size $\lambda$ around the identity. Averaging over this measure is denoted by $\EE$. Associated to any random sequence of matrices are products $\prod_{n=1}^N \mathcal{T}_{\lambda, n}=\mathcal{T}_{\lambda, N}\cdots\mathcal{T}_{\lambda, 1}$ and their assympototics satisfy a $0$-$1$ law characterized by the Lyapunov exponent $\gamma_\lambda$ \cite{BL}. It can be calculated by
\begin{equation}
\gamma_\lambda\;=\;
\lim_{N\to\infty}\frac{1}{N}\;\EE\Bigl[\log(\|\prod_{n=1}^N \mathcal{T}_{\lambda, n} e \|)\Bigr]\,,
\end{equation}
\noindent with some initial condition $e\in\RM^2$, $\|e\|=1$. Furthermore, a central limit theorem \cite{Tut,BL} states that the expression
\begin{equation}
\frac{1}{\sqrt{N}}
\left(\log\bigl(\|\prod_{n=1}^N \mathcal{T}_{\lambda, n}e\|\bigr)\, -\, N\,\gamma_\lambda\right) 
\end{equation}
\noindent converges to a centered Gaussian law with a variance denoted by $\sigma_\lambda$. Both the variance and the Lyapunov exponent are independent of the initial condition $e$. Clearly the distribution of $\Pp_n$ should have a crucial influence on the values of $\gamma_\lambda$ and $\sigma_\lambda$. The following theorem summarizes the main results of the paper.

\begin{theo} 
\label{theo-main}
Let the i.i.d. random matrices $\mathcal{T}_{\lambda, n}$ be of the form {\rm \eqref{eq-matdef}} with the distribution supported on a bounded set. Furthermore, it is supposed that in each of the cases below a certain linear combination of the entries of $\Pp_n$ specified below has a strictly positive variance. 

\begin{itemize}

\item[{\rm (i)}] If the averaged perturbation is elliptic in the sense that $\det(\EE[\mathcal{P}_n])>0$, then there is a constant $C_e$, which  is calculated explicitly from the variances of the entries of $\mathcal{P}_n$ in {\rm Section~\ref{sec-ellfirst}}, such that
$$
\gamma_\lambda
\;=\;C_e\,\lambda^2\,+\,\Oo(\lambda^3)\;,
\qquad
\sigma_\lambda\;=\;C_e\,\lambda^2\,+\,\Oo(\lambda^3)\;.
$$

\item[{\rm (ii)}] If the averaged perturbation is hyperbolic in the sense that $\det(\EE[\mathcal{P}_n])<0$, then with $C_h=\sqrt{-\det(\EM(\Pp_n))}$,
$$
\gamma_\lambda
\;=\;C_h\,\lambda\,+\,\Oo(\lambda^\frac{3}{2})\;,
\qquad
\sigma_\lambda\;=\;\Oo(\lambda^\frac{3}{2})\;.
$$

\item[{\rm (iii)}] If the perturbation is centered $\EE[\mathcal{P}_n]=0$, then there exists constants $C_s$ and $C'_s$ such that
$$
\gamma_\lambda
\;=\;
C_s\,\lambda^2\,+\,\Oo(\lambda^3)\;,
\qquad
\sigma_\lambda\;=\;C'_s\,\lambda^2\,+\,\Oo(\lambda^3)
\;.
$$
\end{itemize}

\end{theo}

\vspace{.2cm}

There have been numerous works in both the physics and the mathematics literature on products of random matrices of the form \eqref{eq-matdef}. One is an influential work of Kappus and Wegner \cite{KW} on so-called band center anomalies for the Lyapunov exponent at the band center of the one-dimensional Anderson model. The term {\it anomaly} reflects that the standard perturbation theory breaks down and that this leads to an enhancement of the Lyapunov exponent. These anomalies and also the band edge were then further studied by Derrida and Gardner \cite{DG}. In the sequel, many rigorous works analyzed the Lyapunov exponent of the Anderson model in these particular citations \cite{BK,CK,SVW,SB,SS1}. In \cite{SS1} any model of the form \eqref{eq-matdef} was called an anomaly and such anomalies were further distinguished to be of first order elliptic if $\det(\EE[\mathcal{P}_n])>0$, of first order hyperbolic if $\det(\EE[\mathcal{P}_n])<0$, and of second order if $\EE[\mathcal{P}_n]=0$. This terminology will be used below and corresponds to the three cases in the theorem. Let us point out that the parabolic first order case $\det(\EE[\mathcal{P}_n])=0$ and $\EE[\mathcal{P}_n]\not=0$ is not covered by the theorem, but can be considered as non-generic.

\vspace{.2cm}

The main novelty of Theorem~\ref{theo-main} are the formulas for the variances. In fact, the Lyapunov exponents in (i) and (iii) were calculated in \cite{SS1} and \cite{SB} respectively, and the case of (ii) was sketched in \cite{SS1,DKS}, but there are actually considerable technical difficulties to make the analysis rigorous, see Section~\ref{sec-hyperbolic}. On the other hand, no other work on variances at anomalies is known to us. The paper \cite{SSS} developed a perturbation theory for the variance in situations where the lowest order matrix $\mathcal{T}_{0, n}$ is a non-trivial rotation rather than the identity. This corresponds to energies away from the band center and the band edges in the Anderson model. To control the variances in Theorem~\ref{theo-main} it is necessary to go far beyond the analysis of \cite{SSS}.

\vspace{.2cm}

In the first order elliptic case of Theorem~\ref{theo-main}(i), the result implies the equality $\sigma_\lambda=\gamma_\lambda+\Oo(\lambda^3)$. Hence the asymptotic distribution of the product of random matrices is described by a single parameter, up to errors of higher order. In the framework of random Schr\"odinger operators this is referred to as single parameter scaling. It was already shown to hold away from anomalies in \cite{SSS}, but only to lowest order in perturbation theory.  This is now extended to the first order elliptic regime. For an Anderson model, this covers energies inside of the band, but close to a band edge (see Section~\ref{sec-examples}). On the other hand, energies outside of the band, but close to a band edge, correspond to a first order hyperbolic anomaly as in Theorem~\ref{theo-main}(ii). In this situation, the Lyapunov exponent is given by its deterministic value, up to fluctuations which are of much smaller order as expected (because the system is a random perturbation around a hyperbolic one, albeit a very weakly hyperbolic one). In particular, there is no single parameter scaling in this regime. We are not able to determine whether $\sigma_\lambda$ is of even lower order than $\Oo(\lambda^{\frac{3}{2}})$. In the centered case of Theorem~\ref{theo-main}(iii), the Lyapunov exponent and variance are of the same order of magnitude, and the formulas in Section~\ref{sec-2order} show that an equality $C_s=C'_s$ is to be considered a coincidence so that single parameter scaling does not hold in this case. 

\vspace{.2cm}

In the final Section~\ref{sec-examples} of the paper, it is shown how Theorem~\ref{theo-main} can be applied to the Anderson model as well as two other models, namely a harmonic chain and the Kronig-Penny model. Both have already been considered in the work of Ishii \cite{Ish}, but there has been continuous and also recent interest in them ({\it e.g.} \cite{AH} and \cite{DKS} respectively). The remainder of the article is mainly devoted to proofs. Apart from perturbative formulas and recursive arguments, the main technical tool is the analysis of singular differential operators.

\vspace{.5cm}

\noindent {\bf Acknowledgements:} We thank the referee for careful reading and several comments that improved the presentation. This work profited from financial support by the DFG.

\newpage

\section{From random products to a random dynamical system}

This section contains some preparatory material, most of which is already contained in \cite{BL,SS1,SSS}, but is needed to fix notations and make this work self-contained.

\subsection{Dynamics on the unit circle}
\label{eq-dynaS1}

For any $P=\binom{a\;\;b}{c\;-a} \in\mbox{sl}(2,\RM)$ one has $P^2=-\det(P)\one_2$ which implies
$$
\exp(\lambda P)
\;=\;
\cosh(\lambda d)\,\one_2\;+\;\frac{\sinh(\lambda d)}{d}\,P
\;,
\qquad
d=\sqrt{-\det(P)}\in\CM
\;.
$$
A dynamics $\theta\in\SM^1_\pi\mapsto S_P(\theta)\in\SM^1_\pi$ on $\SM^1_\pi=[0,\pi)$ is induced by 
$$
e_{S_{\lambda P}(\theta)}
\;=\;
\frac{\exp(\lambda P)e_\theta}{\|\exp(\lambda P)e_\theta\|}
\;,
\qquad
e_{\theta} =
\begin{pmatrix}
\cos(\theta) \\ \sin(\theta)
\end{pmatrix}
\;.
$$
This dynamics can readily be calculated in terms of harmonics in $\theta$, namely with $v=\binom{ 1}{\imath}$
$$
e^{2\imath S_{\lambda P}(\theta)}
\;=\;
\frac{v^* \exp(\lambda P) e_\theta}{\overline{v}^* \exp(\lambda P) e_\theta}
\;=\;
\frac{\cosh(\lambda d)e^{\imath \theta}\,+\,\frac{\sinh(\lambda d)}{d}\,v^*Pe_\theta}{
\cosh(\lambda d)e^{-\imath \theta}\,+\,\frac{\sinh(\lambda d)}{d}\,\overline{v}^*Pe_\theta}
\;,
$$
so that in terms of the entries of $P$
$$
e^{2\imath (S_{\lambda P}(\theta)-\theta)}
\;=\;
\frac{\cosh(\lambda d)\,+\,\frac{\sinh(\lambda d)}{d}\bigl(a e^{-2\imath \theta}
+\frac{\imath}{2}(c-b)+\frac{\imath}{2}(c+b)e^{-2\imath \theta}\bigr)
}{
\cosh(\lambda d)\,+\,\frac{\sinh(\lambda d)}{d}\bigl(a e^{2\imath \theta}
-\frac{\imath}{2}(c-b)-\frac{\imath}{2}(c+b)e^{2\imath \theta}\bigr)
}
\;.
$$
Expanding in $\lambda$ therefore leads to
$$
S_{\lambda P}(\theta)
\;=\;
\theta
\;+\;\lambda \,p(\theta)\;+\;\frac{1}{2}\,\lambda^2\,p(\theta)\,\partial_\theta p(\theta)\;+\;
\Oo(\lambda^3)
\;,
$$
where, still for $P=\begin{pmatrix}a & b \\ c & -a \end{pmatrix}$,
\begin{equation}
\label{eq-pdef}
p(\theta)
\; =\;
-a \sin(2\theta)\,+\,\tfrac{1}{2}(c-b)\,+\,\tfrac{1}{2}(c+b)\cos(2\theta) \;=\; -a \sin(2\theta)\,-\,b\,\sin^2(\theta)\,+\,c\,\cos^2(\theta)
\;.
\end{equation}

\vspace{.2cm}

All the above applies directly to $\mathcal{T}_{\lambda, n}$. However, in order to deal with the various types of anomalies, it will be advantageous to make an appropriate basis change $M\in\mbox{Gl}(2, \RM)$ from $\mathcal{T}_{\lambda, n}$ to
$$
T_{\lambda, n}
\;=\;
M \mathcal{T}_{\lambda, n} M^{-1}\;=\;\exp(\lambda P_n + \lambda^2 Q_{\lambda,n})
\;,
$$
with $P_n = M\mathcal{P}_n M^{-1}$ and $Q_{\lambda,n} = M\mathcal{Q}_{\lambda,n} M^{-1}$ respectively. The adequate choice of $M$ will be made for each type of anomaly further below. Let us denote the dynamics associated with $T_{\lambda,n}$ by $S_{\lambda,n}=S_{\lambda,P_n+\lambda Q_{\lambda,n}}$. According to the above,
\begin{equation}
\label{eq-dynexpan}
S_{\lambda,n}(\theta)
\;=\;
\theta
\;+\;\lambda \,p_n(\theta)\;+\;\lambda^2\,q_n(\theta)\;+\;\frac{1}{2}\,\lambda^2\,p_n(\theta)\,\partial_\theta p_n(\theta)\;+\;
\Oo(\lambda^3)
\;,
\end{equation}
where $p_n$ and $q_n$ are defined by \eqref{eq-pdef} from the entries of $P_n$ and $Q_{0,n}$ respectively. Now given any sequence $T_{\lambda,n}$ or equivalently $(P_n,Q_{\lambda,n})_{n\geq 1}$ and any initial condition $\theta_0\in\SM^1_\pi$, one obtains a sequence
$$
\theta_{n+1}
\;=\;
S_{\lambda,n}(\theta_n)
\;.
$$
If $P_n$ and $Q_{0,n}$ are random, this provides a random dynamical system on the circle $\SM^1_\pi$.

\subsection{Invariant measure}

There is an invariant measure $\nu_\lambda$ on $\SM^1_\pi$, corresponding to the action of the family $(T_{\lambda, n})_{n\in\NM}$, which is given by
$$
 \int_0^{\pi} f(\theta)\,\nu_\lambda(d\theta)\;=\;\EE\int_0^{\pi}f(S_{\lambda, n}(\theta))\,\nu_\lambda(d\theta)\, , \qquad f\in C(\SM^1_\pi)\,.
$$
From the Furstenberg theorem \cite{BL} follows that the invariant measure is unique and H\"older continous, provided the Lyapunov exponent of the associated system is positive. This actually holds in all cases considered here. Of interest will be the following ergodic sums
$$
I_{\lambda,N}(f)\;=\;\frac{1}{N}\;\,\EE \sum_{n=0}^{N-1} f(\theta_n)\,,
$$
and especially in their limit
$$
I_\lambda(f)\;=\;\lim_{N\to\infty} I_{\lambda,N}(f)\,.
$$
This limit can also be expressed in the terms of the invariant measure $\nu_\lambda$ for which several notations are used:
$$
 I_\lambda(f)\;=\;\int_0^{\pi} f(\theta)\,\nu_\lambda(d\theta)
\;=\;
\nu_\lambda(f)
\;=\;
\EM_{\nu_\lambda}[f]
\,.
$$
As in \cite{SSS} the calculation of the variance $\sigma_\lambda$ is based on a control of the correlation sum 
$$
J_\lambda(f)\;=\;\EE_1\left[\sum_{n=1}^\infty \bigl(f(\theta_n) - \nu_\lambda(f)\bigr)\right]\,.
$$
Here $J_\lambda(f)$ depends on the initial condition $\theta_0$ and $\EE_n$ denotes the expectation over all
$T_{\lambda,m}$ with $m\geq n$. Hence $\EE_1=\EE$. It is shown in \cite{BL} that positivity of the Lyapunov exponent implies that $J_\lambda(f)$ is finite for H\"older continuous functions. Let us point out that $f\mapsto J_\lambda(f)$ is linear and that constant functions are in the kernel of this map.  A quantitative bound is given in Proposition~\ref{a-priori} below.

\subsection{The Lyapunov exponent and the variance}

Let us introduce a family $(g_n)_{n\geq 1}$ of random variables by 
$$
 g_n\;=\;\frac{1}{2}\log(\|T_{\lambda, n} e_{\theta_{n-1}} \|^2)\,.
$$
The Lyapunov exponent can then be defined as
\begin{equation}
\gamma_\lambda
\;=\;
\lim_{N\to\infty}
\frac{1}{N}\;
\EE\,\sum_{n=1}^N\,g_n
\;=\;
I_\lambda(g)
\label{eq-gamma1}
\mbox{ , }
\end{equation}
where $g:\SM^1_\pi\rightarrow\RM$ is given by $g(\theta_{n-1})=\EE_n[g_n]$, while the variance is
\begin{eqnarray}
\sigma_\lambda
\!\! & = & \!\!
\lim_{N\to\infty}\frac{1}{N}\;
\sum_{n,k=1}^N\,\EE_{\nu_\lambda}\,
\Big[(g_n-\gamma_\lambda)
\,(g_k-\gamma_\lambda)
\Big]
\nonumber
\\
\!\! & = &\!\!
\lim_{N\to\infty}\frac{1}{N}\;
\sum_{n=1}^N\;
\EE_{\nu_\lambda}\!
\left[
g_n^2-\gamma_\lambda^2
+2\sum_{m=1}^{N-n}
\Big(g_n g_{n+m}-\gamma^2_\lambda\Big)
\right]
\;.
\nonumber
\end{eqnarray}
The positivity of the Lyapunov exponent implies again \cite{BL} that the sum over $m$ is convergent even if
$N\to\infty$. Furthermore,  $\EE_n(g_{n+m})$ converges exponentially
fast to $\gamma_\lambda$ as $m\to\infty$ \cite{BL,SSS} and the summands in the sum over $n$ 
converge in expectation so that (with integration w.r.t. $\nu_\lambda$ over $\theta_0$)
\begin{align}
\sigma_\lambda
&=\;
\EE_{\nu_\lambda}\EE
\left[
g_1^2-\gamma^2_\lambda
+2\sum_{m=2}^{\infty}
\left(g_1g_m-\gamma^2_\lambda\right)
\right]\;=\;
\EE_{\nu_\lambda}\EE
\left[
g_1^2-\gamma^2_\lambda
+2\,g_1\,\EE_2\,\sum_{m=2}^{\infty}
\left(g_m-\gamma_\lambda)\right)
\right]
\nonumber
\\ \label{eq-sigmalambda}
\\
&=\;
\EE_{\nu_\lambda}\EE
\left[
g_1^2(\theta_0)-\gamma^2_\lambda
+2\,g_1(\theta_0)J_\lambda(g)(\theta_1)
\right]
\;=\;\EE_{\nu_\lambda}
\left[
g^2-\gamma^2_\lambda
+2\,\EE(g_1 J_\lambda(g)\circ S_{\lambda, 1})\right]
\;.\nonumber
\end{align}
%

\subsection{Basic perturbative formulas and estimates}

Replacing the expansion
$$
\|T_{\lambda, n} e_\theta \|^2
\;=\;
1+2\lambda e_\theta^*P_n e_\theta\;+\;
\lambda^2(2 e_\theta^*Q_n e_\theta + e_\theta^*P_n^* P_n e_\theta -\det(P_n))\,+\,\Oo(\lambda^3)\,,
$$
into the definition of $g_n$ leads to 
$$
g_n
\;=\;
\lambda e_{\theta_{n-1}}^*P_n e_{\theta_{n-1}}\;+\;
\lambda^2(e_{\theta_{n-1}}^*Q_n e_{\theta_{n-1}} +\frac{1}{2} e_{\theta_{n-1}}^*P_n^* P_n e_{\theta_{n-1}} -\frac{1}{2}\det(P_n) - ( e_{\theta_{n-1}}^*P_n e_{\theta_{n-1}})^2 )\,,
$$
up to terms of order $\Oo(\lambda^3)$. Using
$$ 
e_\theta^*Ae_\theta\;=\;\frac{\Tr(A)}{2} + \frac{a-d}{2}\cos(2\theta) + \frac{b+c}{2}\sin(2\theta)\,,
\qquad
A\,=\,\begin{pmatrix}a & b \\c & d \end{pmatrix}
\;,
$$
one can rewrite $g_n$ in terms of the entires of $P_n=
\begin{pmatrix}
p_{1,n} & p_{2,n}\\
p_{3,n} & -p_{1,n}               
\end{pmatrix}$ 
and $Q_n=
\begin{pmatrix}
q_{1,n} & q_{2,n}\\
q_{3,n} & -q_{1,n}               
\end{pmatrix}$:
\begin{align}
g_n
\;= & \lambda \Bigl(p_{1,n}\cos(2\theta_{n-1}) + \frac{1}{2}(p_{2,n} + p_{3,n})\sin(2\theta_{n-1}) \Bigr) 
\;+\;\lambda^2\Bigl(\frac{1}{8} [4p_{1,n}^2 + (p_{2,n} + p_{3,n})^2] 
\nonumber
\\
& \;+\;[q_1 + \frac{1}{4}(p_{3,n}^2 - p_{2,n}^2)]\cos(2\theta_{n-1}) + \frac{1}{2}[q_{2,n}+q_{3,n} + p_{1,n}(p_{2,n} - p_{3,n})]\sin(2\theta_{n-1}) 
\label{gamma_n}
\\ 
& \; - \,\frac{1}{8}[4p_{1,n}^2-(p_{2,n}+p_{3,n})^2]\cos(4\theta_{n-1}) - \frac{1}{2}[p_{1,n}(p_{2,n}+p_{3,n})]\sin(4\theta_{n-1})\Bigr) \;+ \;\Oo(\lambda^3)\,.
\nonumber
\end{align}
Using these formulas, the following quantitative {\it a priori} estimate on $J_\lambda(f)$ is proved in \cite{SSS}.

\begin{proposi}
\label{a-priori}
There is a constant $C$ such that for any H\"older continous function $f$ with H\"older norm $\|f\|_\alpha$, 
$$
J_\lambda(f) 
\;\leq\; 
\frac{C\,\|f\|_\alpha}{\lambda\gamma_\lambda}
 \;.
$$
\end{proposi}

\vspace{.2cm}


\section{Elliptic first order anomaly}
\label{sec-ellfirst}

This section considers the elliptic first order anomaly and proves item (i) of Theorem~\ref{theo-main}. As $\det(\EE[\Pp_n])>0$, there is a basis transformation by $M$ such that the matrix $\EE[P_n]$ is the generator of a rotation:
\begin{equation}
\label{elliptic}
P_n\;=\;
\begin{pmatrix}
0 & -\eta \\
\eta & 0
\end{pmatrix}
\;+\; \widetilde{P}_n
\;=\;
\begin{pmatrix}
\widetilde{p}_{1,n} & \widetilde{p}_{2,n} - \eta \\
\widetilde{p}_{3,n} + \eta & -\widetilde{p}_{1,n}
\end{pmatrix}
\,,
\end{equation}
where $\eta=\sqrt{\det\EE[P_n]}$ and $\widetilde{P}_n$ is centered. Thus to lowest order the dynamics is a deterministic rotation perturbed by random (centered) fluctuations:
\begin{equation}
\label{eq-Sexpa}
S_{\lambda, n}(\theta)\;=\;\theta+\lambda\eta+\lambda
\left(\frac{1}{2}(\widetilde{p}_{2,n}-\widetilde{p}_{3, n}) + \frac{1}{2}(\widetilde{p}_{2,n}+\widetilde{p}_{3, n})\cos(2\theta)-\widetilde{p}_{1, n}\sin(2\theta)\right)
\;+\;
\Oo(\lambda^2)\,.
\end{equation}

\begin{proposi} {\rm \cite{SS1}}
Up to corrections, the invariant measure at an elliptic first order anomaly is given for $f\in C^2(\SM^1_\pi)$ by the Lebesgue measure:
$$
I_\lambda(f)\;=\;\frac{1}{\pi}\int_0^{\pi}f(\theta)d\theta \;+\; \Oo(\lambda)\,.
$$
\end{proposi}

\noindent {\bf Proof.} Expanding $f\in C^2(\SM^1_\pi)$ in Taylor series and taking the expectation provides
$$
\EE_{n+1}[f(S_{\lambda,n}(\theta_n))]\;=\;f(\theta_n)\,+\,\lambda\,\eta\, f'(\theta_n) + \Oo(\lambda^2)\,.
$$
By summing this up one obtains
$$
I_{\lambda,N}(f)\;=\;I_{\lambda,N}(f) \,+\,\lambda\,\eta \,I_{\lambda,N}(f') \,+\, \Oo(\lambda^2, N^{-1})\;,
$$
\noindent and since $f-\frac{1}{\pi}\int_0^{\pi}f(\theta)d\theta$ has an antiderivative, the statement follows.
\hfill $\Box$

\vspace{.2cm}

In particular, $I_\lambda(\cos(2\cdot))=I_\lambda(\sin(2\cdot))=\Oo(\lambda)$. This allows to calculate the Lyapunov exponent.
Using \eqref{elliptic}  one gets from \eqref{gamma_n}
\begin{eqnarray*}
g_n
& = &
\lambda\left( \widetilde{p}_{1,n}\cos(2\theta_{n-1}) + \frac{1}{2}( \widetilde{p}_{2,n} +  \widetilde{p}_{3,n})\sin(2\theta_{n-1})\right) 
\;+\;\lambda^2\Bigl(\frac{1}{8} [4 \widetilde{p}_{1,n}^2 + ( \widetilde{p}_{2,n} +  \widetilde{p}_{3,n})^2] \\
& &
\;+\;[q_1 + \frac{1}{4}( \widetilde{p}_{3,n}^2 -  \widetilde{p}_{2,n}^2)-\frac{1}{2}\eta( \widetilde{p}_{2,n} +  \widetilde{p}_{3,n})]\cos(2\theta_{n-1})
\\
& & \, +\, \frac{1}{2}[q_{2,n}+q_{3,n} +  \widetilde{p}_{1,n}( \widetilde{p}_{2,n} - \widetilde{p}_{3,n}-2\eta)]\sin(2\theta_{n-1}) \\
& & - \frac{1}{8}[4 \widetilde{p}_{1,n}^2-( \widetilde{p}_{2,n}+ \widetilde{p}_{3,n})^2]\cos(4\theta_{n-1}) - \frac{1}{2}[ \widetilde{p}_{1,n}( \widetilde{p}_{2,n}+ \widetilde{p}_{3,n})]\sin(4\theta_{n-1})\Bigr) \;+\; \Oo(\lambda^3)\,,
\end{eqnarray*}
so that the Lyapunov exponent is
\begin{equation}
\label{gamma_ell}
 \gamma_\lambda\;=\;\frac{\lambda^2}{8}\,(4\mbox{Var} (\widetilde{p}_1) + \mbox{Var} (\widetilde{p}_2 + \widetilde{p}_3)) \;+\;\Oo(\lambda^3) \,.
\end{equation}
\noindent Here we omitted the indexation with $n$, since the $P_n$'s are distributed identically. This formula also already appeared in \cite{SS1}. The calculation of the variance uses the following lemma.

\begin{lemma}
\label{lem-Jell}
Let $f\in C^{3+\alpha}(\SM^1_\pi)$ with $\alpha>0$. Choose $c\in\RM$ such that the function $f-c$ has an antiderivative $F\in C^{4+\alpha}(\SM^1_\pi)$. Then, independent of the choice of $F$,
$$
J_\lambda(f)\;=\;-\,\frac{1}{\lambda\,\eta}\,
\Big(
F(\theta_0)\,-\,{\nu_\lambda}(F)
\Big)
\,+\,\Oo(1)\,.  
$$
\end{lemma}

\noindent {\bf Proof.}  Let us begin by noting that Proposition~\ref{a-priori} combined with $\gamma_\lambda=\Oo(\lambda^2)$ as given by (\ref{gamma_ell}) shows that $J_\lambda(f)=\Oo(\lambda^{-3})$ because $f$ is  a H\"older continuous function. Furthermore, one has $J_\lambda(f-c)=J_\lambda(f)$ so that it is possible to suppose $c=0$ and $\int^{\pi}_0 f(\theta)\,d\theta=0$ from now on. Next let us recall the Taylor formula
$$
F(\theta+\epsilon)
\;=\;
F(\theta)\,+\,\epsilon\,F'(\theta)\,+\,\frac{1}{2}\,\epsilon^2\,F''(\theta+\widetilde{\epsilon}\,)
$$
where $0\leq \widetilde{\epsilon}/\epsilon\leq 1$. Replacing $\theta=\theta_n$ and $\epsilon=\theta_{n+1}-\theta_n=S_{n,\lambda}(\theta_n)-\theta_n$ and taking the average of \eqref{eq-Sexpa} leads to
$$
\EM_{n}\,
F(\theta_{n+1})
\;=\;
\EM_{n}\,
F(S_{\lambda,n}(\theta_{n}))
\;=\;
F(\theta_n)\,+\,\lambda\,\eta\,F'(\theta_n)\;+\:\lambda^2\,r(\theta_n)
\;,
$$
where $r$ is some H\"older continuous function because $f\in C^{1+\alpha}(\SM^1_\pi)$. Replacing this into $J_\lambda$ and separating the first summand, one has
$$
J_\lambda(F)\;=\;
\Big(
F(\theta_0)\,-\,\EE_{\nu_\lambda}(F)
\Big)
\;+ \;J_\lambda(F)\, +\, \lambda\, \eta \,J_\lambda(F') \,+ \,\lambda^2\,J_\lambda(r)\;.
$$
Therefore, for $f=F'$,
\begin{equation}
J_\lambda(f)
\;=\;
-\;\frac{1}{\eta\,\lambda}\,
\Big(
F(\theta_0)\,-\,{\nu_\lambda}(F)
\,+ \,\lambda^2\,J_\lambda(r)
\Big)
\label{eq-solv}
\end{equation}
As $J_\lambda(r)=\Oo(\lambda^{-3})$ by the a priori estimate, this implies $J_\lambda(f) = \Oo(\lambda^{-2})$. Up to now only $f\in C^{1+\alpha}(\SM^1_\pi)$ was used. As $f\in C^{2+\alpha}(\SM^1_\pi)$, one has $r\in C^{1+\alpha}(\SM^1_\pi)$ and hence also $J_\lambda(r) = \Oo(\lambda^{-2})$. This implies $J_\lambda(f) = \Oo(\lambda^{-2})$ for $f\in C^{2+\alpha}(\SM^1_\pi)$. One further iteration is needed to calculate the value of $J_\lambda(f)$. As $f\in C^{3+\alpha}(\SM^1_\pi)$, one has $r\in C^{2+\alpha}(\SM^1_\pi)$ so that $J_\lambda(r) = \Oo(\lambda^{-1})$. Replacing this in \eqref{eq-solv} concludes the proof.
\hfill
$\Box$

\vspace{.2cm}

Lemma~\ref{lem-Jell} implies $J_\lambda(\cos(j\cdot))=-\frac{1}{j\eta\lambda}\sin(j\theta_0) + \Oo(1)$ and $J_\lambda(\sin(j\cdot))=\frac{1}{j\eta\lambda}\cos(j\theta_0) + \Oo(1)$. For the calculation of the variance the index $n$ is again suppressed:
\begin{eqnarray*}
J_\lambda(g)&=&\frac{\lambda}{2\eta}\Bigl(-\EE[q_1 + \frac{1}{4}(\widetilde{p}_3^2 - \widetilde{p}_2^2)]\sin(2\theta_0) + \frac{1}{2}\EE[q_2+q_3 + \widetilde{p}_1(\widetilde{p}_2 - \widetilde{p}_3)]\cos(2\theta_0) \\
& & + \frac{1}{16}\EE[4\widetilde{p}_1^2-(\widetilde{p}_2+\widetilde{p}_3)^2]\sin(4\theta_0) - \frac{1}{4}\EE[\widetilde{p}_1(\widetilde{p}_2+\widetilde{p}_3)]\cos(4\theta_0) \Bigr) + \Oo(\lambda^2)
\end{eqnarray*}
Replacing into \eqref{eq-sigmalambda} shows that $\sigma_\lambda=\gamma_\lambda+ \Oo(\lambda^3)$ with $\gamma_\lambda$ given by \eqref{gamma_ell}.

\section{Hyperbolic first order anomaly}
\label{sec-hyperbolic}

A hyperbolic first order anomaly corresponds to item (ii) of Theorem~\ref{theo-main}. Now it is possible to choose a basis change $M$ such that
\begin{equation}
\label{eq-hypcoord}
P_n\;=\;\begin{pmatrix}p_{n,1} & p_{n,2} \\ p_{n,3} & -p_{n,1} \end{pmatrix}
\;,
\qquad
\EE(p_{n,2})=\EE(p_{n,3})=0\;,
\;\;\;
\eta=\EE(p_{n,1})>0
\;.
\end{equation}
Then the dynamics is to lowest order (higher orders are irrelevant and thus suppressed in the following)
\begin{equation}
\label{eq-hypcoordS}
S_{\lambda, n}(\theta)\;=\;\theta+\lambda
\left(
-\,p_{n,1}\sin(2\theta)
\,-\,
p_{n,2}\sin^2(\theta)
\,+\,
p_{n,3}\cos^2(\theta)
\right)
\;+\;
\Oo(\lambda^2)\,.
\end{equation}
Hence the averaged dynamics is simply $\EM \,S_{\lambda, n}(\theta)=\theta-\lambda\,\EM(p_{n,1})\sin(2\theta)+\Oo(\lambda^2)$ which has $\theta=0$ as a stable  and $\theta=\frac{\pi}{2}$ as an unstable fixed point, and no other fixed points. Hence, under the averaged dynamics any initial point $\theta_0$ ultimately reaches the stable fixed point $0$, except if $\theta_0=\frac{\pi}{2}$ is the unstable fixed point. The first aim is to show that this behavior is stable in the sense that the invariant measure $\nu_\lambda$ converges to a Dirac peak on the stable fixed point as $\lambda\to 0$. This is actually a delicate endeavor as will become apparent shortly. Once $\nu_\lambda$ is analyzed, Lyapunov exponent and variance can be deduced in Section~\ref{sec-hyplyap}.

\subsection{Random dynamics on $\SM^1$ with two fixed points}
\label{sec-inv2FP}

Let us consider for a  function $F\in C^2(\SM^1_\pi)$ the Taylor expansion
$$
F(S_{\lambda, n}(\theta))\;=\;F(\theta)\, +\, \lambda\, p_n(\theta)F'(\theta)\,+\,\Oo(\lambda^2\|F\|_{C^2})\,.
$$
where $p_n(\theta)=-p_{n,1}\sin(2\theta)-p_{n,2}\sin^2(\theta)+p_{n,3}\cos^2(\theta)$. Thus the associated ergodic sums satisfy
$$
I_{\lambda,N}(F)\;=\;\frac{1}{N}\sum_{n=0}^{N-1}\EE[F(\theta_n)]\;=\;I_{\lambda,N}(F)\,+\, \lambda \,I_{\lambda,N}(\EE[p_n]F')\,+\,\Oo(\lambda^2\|F\|_{C^2}, N^{-1})\,, 
$$
so that  for every function $f=\Mm F$ in the range of the first order differential operator $\Mm:\,C^1(\SM_\pi^1)\rightarrow C(\SM_\pi^1)$, defined by 
\begin{equation}
\label{eq-Ldef}
(\Mm F)(\theta)
\;=\;
\eta\,\sin(2\theta)\,F'(\theta)
\;,
\end{equation}
one has $I_{\lambda,N}(f)=\Oo(\lambda\|F\|_{C^2}, N^{-1})$. Thus, upon integration w.r.t. the invariant probability measure $\nu_\lambda$, one has $\nu_\lambda(f)=\Oo(\lambda\|F\|_{C^2})$.  Every function in $\Ran(\Mm)$ has zeroes of order at least $1$ in both fixed points, but $\Ran(\Mm)$ contains sufficiently many functions to conclude that the weight of $\nu_\lambda$ is of order $\Oo(\lambda^{\frac{1}{2}})$ outside of $\lambda^{\frac{1}{4}}$-neighborhoods of the both fixed points. Indeed, let us choose $F(\theta)=-\cos(2\theta)/2\eta$ so that $f(\theta)=(\Mm F)(\theta)=\sin^2(2\theta)$. Thus $I_\lambda(\sin^2(2\cdot))=\Oo(\lambda)$ so that, using the indicator function $\chi_A$ on $A=\{\theta\in\SM^1_\pi\,|\,\sin^2(2\theta)>\lambda^{\frac{1}{2}}\}$ which is bounded by $\chi_A(\theta)\leq \lambda^{-\frac{1}{2}}\sin^2(2\theta)$, one deduces $\nu_\lambda(A)=\Oo(\lambda^{\frac{1}{2}})$. 

\vspace{.2cm}

\noindent {\bf Remark} Of course, it is possible to implement this line of argument for many other functions. For example, taking $F(\theta)=\sin(2\theta)$ one obtains $I_\lambda(\sin(4\cdot))=\Oo(\lambda)$. Analogously, with $F(\theta)=\sin((2+4k)\theta)$ for $k\in\NM$, one obtains inductively $I_\lambda(\sin(4k\cdot))=\Oo(\lambda)$. Similarly, one obtains  $I_\lambda(\cos(4k\cdot))=1 + \Oo(\lambda)$ for $k\in\NM$. Furthermore, using directly the dynamics \eqref{eq-hypcoordS}
\begin{equation}
\label{eq-dyndirectexp}
 \frac{1}{N}\,\sum_{n=0}^{N-1}\EE\,\theta_n\;=\;\frac{1}{N}\,\sum_{n=0}^{N-2}\EE\,\theta_n - \lambda\,\eta\,\frac{1}{N}\sum_{n=0}^{N-2}\EM\,\sin(2\theta_n) \,+\, \Oo(\lambda^2, N^{-1})\,,
\end{equation}
so that $I_\lambda(\sin(2\cdot))=\Oo(\lambda)$ and, by the above procedure, $I_\lambda(\sin(2k\cdot))=\Oo(\lambda)$ for all $ k\in\NM$.
\hfill $\diamond$

\vspace{.2cm}

However, the above strategy does not allow to say what the distribution of $\nu_\lambda$-weight of balls centered at the two fixed points $0$ and $\frac{\pi}{2}$ actually is. For example, one cannot evaluate $I_\lambda(\cos(2\cdot))$ by this method (its value is then equal to $I_\lambda(\cos((2+4k)\cdot))$ for $k\in\NM$) because the function $\theta\mapsto\cos(2\theta)$ takes different values at the two fixed points. To go further it is necessary to zoom the dynamics into the unstable fixed point $\frac{\pi}{2}$ and show that the fluctuations drive the random dynamics out of the neighborhood of this point.  This will be done by a change of variables induced by a special M\"obius transformation, in a manner explained in \cite{SS1}:
$$
\begin{pmatrix} \lambda^{-\frac{1}{2}} & 0 \\ 0 & 1 \end{pmatrix}
\,\exp\left( \lambda\,\begin{pmatrix}p_1 & p_2 \\ p_3 & -p_1 \end{pmatrix}\right)\,
\begin{pmatrix}\lambda^{\frac{1}{2}} & 0 \\ 0 & 1 \end{pmatrix}
\;=\;
\exp\left( \lambda^{\frac{1}{2}}\,\begin{pmatrix} 0 & p_2 \\ 0 & 0 \end{pmatrix}
+\lambda\begin{pmatrix} p_1 & 0 \\ 0 & -p_1 \end{pmatrix}
+
\Oo(\lambda^{\frac{3}{2}})
\right)
\;,
$$
where the abbreviation $p_{n,j}=p_j$ was used. The dynamics induced by the r.h.s. will be denoted by $\hat{S}_{\lambda, n}:\SM^1_\pi\to\SM^1_\pi$ and the associated random dynamics by $(\hat{\theta}_n)_{n\geq 0}$. All objects in this rescaled representation will carry a hat in the sequel. One finds from \eqref{eq-dynexpan}
\begin{equation}
\label{eq-Sexpan}
 \hat{S}_{\lambda, n}(\hat{\theta})\;=\; \hat{\theta}\,-\,\lambda^{\frac{1}{2}} p_{n,2}\sin^2(\hat{\theta})\, +\, \lambda \big(\tfrac{1}{2}\,p_{n,2}^2\,\sin^2(\hat{\theta})-p_{n,1}\big)\sin(2\hat{\theta})
 \,+\, 
 \Oo\big(\lambda^{\frac{3}{2}}\big) 
\;.
\end{equation}
Now the lowest order term (in $\lambda^{\frac{1}{2}}$) is centered, and  the first term with non-vanishing mean is precisely of the order (here $\lambda$) given by the square of the lowest order term.  Hence one is a situation where the techniques of \cite{SS1}, in particular Proposition 3(vii), can be applied. However, we present a self-contained argument here in order to show that almost all the $\nu_\lambda$-weight lies in a neighborhood of the stable fixed point $0$.

\begin{proposi}
\label{prop-two-fp}
Suppose $\EM[p_2^2]>0$. Let $\chi_B$ be the indicator function on $B=\{\theta\in\SM^1_\pi\,|\,|\theta|>\lambda^{\frac{1}{4}}\}$. Then
$$
\nu_\lambda(B)\;=\;\Oo(\lambda^{\frac{1}{2}})
\;.
$$
\end{proposi}

\noindent {\bf Proof:} Underlying the above basis change to the rescaled dynamics \eqref{eq-Sexpan} is the diffeomorphism $t_\lambda:\SM_\pi^1\rightarrow \SM_\pi^1$ defined by
$$
t_\lambda(\hat{\theta})\;=\;\tan^{-1}(\lambda^{-\frac{1}{2}}\tan(\hat{\theta}))
\;,
$$
It zooms into the unstable fixed point $\frac{\pi}{2}$ and keeps $0$ conserved. Then $\hat{S}_{\lambda, n}= 
t_\lambda^{-1}\circ S_{\lambda, n} \circ t_\lambda$ and one is led to study the random rescaled dynamics $(\hat{\theta}_n)_{n\geq 0}$ given by $\hat{\theta}_n=t_\lambda^{-1}({\theta}_n)$. The aim is to control the Birkhoff sums $\hat{I}_N(\hat{f}_\lambda)$ of the rescaled dynamics for the functions $\hat{f}_\lambda= f\circ t_\lambda$, which are readily connected to the original ones via
$$
I_{\lambda,N}(f)\;=\;\frac{1}{N}\sum_{n=0}^{N-1}\mathbb{E}\,f(\theta_n)\;=\;\frac{1}{N}\sum_{n=0}^{N-1}\mathbb{E}\,f(t_\lambda(\hat{\theta}_n))\;=\;\hat{I}_{\lambda,N}(\hat{f}_\lambda)\,.
$$
With the aim of controlling such Birkhoff sums let us again expand for a given function $\hat{F}\in C^2(\SM^1_\pi)$ using \eqref{eq-Sexpan} and $\EM(p_2)=0$: 
$$
\EM\,\hat{F}(\hat{S}_{\lambda, n}(\hat{\theta})) 
\;=\; 
\hat{F}(\hat{\theta})
 \,+\,\lambda\,\frac{\EM(p^2_{2})}{2}
\Bigl(\sin^4(\hat{\theta})\hat{F}''(\hat{\theta})+ \big(\sin^2(\hat{\theta})-2\,k\big)
\sin(2\hat{\theta})\hat{F}'(\hat{\theta})\Bigr)
\,+\,\Oo(\lambda^{\frac{3}{2}}\|\hat{F}\|_{C^2}) \;,
 $$
where $k=\EM(p_{1})/\EM(p^2_{2})>0$.
Defining a first order differential operator $\hat{\mathcal{L}}:C^1(\SM^1_\pi)\to C(\SM^1_\pi)$ by 
\begin{equation}
\label{eq-Lhatdef}
\bigl(\hat{\mathcal{L}}\,\hat{g}\bigr)(\hat{\theta})\;=\;\sin^4(\hat{\theta})
\,\hat{g}'(\hat{\theta})+ \big(\sin^2(\hat{\theta})-2\,k\big)\sin(2\hat{\theta})\,\hat{g}(\hat{\theta})\,,
\end{equation}
one concludes that the Birkhoff sums satisfy
$$
\hat{I}_{\lambda,N}(\hat{F})\;=\;\hat{I}_{\lambda,N}(\hat{F})\,+\,\lambda\,\frac{\EM(p^2_{2})}{2} \,\hat{I}_{\lambda,N}(\hat{\mathcal{L}}\,\hat{F}')
\,+\,
\Oo(\lambda^{\frac{3}{2}}\|\hat{F}\|_{C^2} ,N^{-1})\,. 
$$
Therefore the function $\hat{\rho}=\hat{\mathcal{L}}\,\hat{F}'$ satisfies
\begin{equation}
\label{eq-hyp}
\hat{I}_{\lambda,N}(\hat{\rho})\;=\;
\Oo\big(\lambda^{\frac{1}{2}}\|\hat{F}\|_{C^2},N^{-1}\big)\,.
\end{equation}
Let us choose in \eqref{eq-hyp}
$$
\hat{\rho}(\hat{\theta})
\;=\;
\exp\bigl(- \,k\,\lambda^{\frac{1}{2}}\,\cot^2(\hat{\theta}) \bigr)
\;,
$$
then solve $\hat{\rho}=\hat{\mathcal{L}}\,\partial_{\hat{\theta}}\hat{F}$ for $\hat{F}$ and show that this solution satisfies $\|\hat{F}\|_{C^2}\leq C$ uniformly in $\lambda$ (note that $\hat{F}$ depends on $\lambda$ via $\hat{\rho}$). Hence one can first solve the first order equation $\hat{\mathcal{L}}\hat{G}=\hat \rho$ for $\hat{G}$. There is actually a two-parameter family of solutions $\hat{G}$ in the neighborhood of the singularity $0$ due to Proposition~3(vii) in \cite{SS1} (because the order of singularities and the signs of the coefficients are precisely as required there).  The first parameter is fixed by requiring $\hat{G}$ to be continuous on all $\SM^1_\pi$ (periodicity) and the second one by adding an adequate multiple of  the (smooth non-negative) ground state $\hat{G}_0$ satisfying $\hat{\mathcal{L}}\hat{G}_0=0$, explicitly given by
$$
\hat{G}_0(\hat{\theta})\;=\;\frac{\exp\bigl(- \,k\,\sin^{-2}(\hat{\theta}) \bigr)}{\sin^2(\hat{\theta})}
\;.
$$
This assures that $\hat{G}$ has vanishing integral and therefore has an antiderivative $\hat{F}$. This explicit solution $\hat{G}=\hat{F}'$ of $\hat{\mathcal{L}}\hat{G}=\hat{\rho}$ is (as in \cite{SS1})
\begin{equation}
\label{eq-F'}
\hat{F}'(\hat{\theta})\;=\;\frac{\exp\bigl(- \,k\,\sin^{-2}(\hat{\theta}) \bigr)}{\sin^2(\hat{\theta})}\left( C_1\,+\,
\int_{C_2}^{\hat{\theta}}\hat{\rho}(\hat s)\;
\frac{\exp\bigl(k\,\sin^{-2}(\hat{s}) \bigr)}{\sin^2(\hat s)}\,d\hat s\right)\,.
\end{equation}
The function $\hat{F}'$ and its derivative $\hat{F}''$ are obviously regular on an interval $I\subset\SM^1_\pi$ not containing $\hat{\theta}=0$ with bounds on $I$ that are uniform in $\lambda$ (because $\hat{\rho}\leq 1$ uniformly in $\lambda$). More delicate is the behavior at $\hat{\theta}=0$ because the integrand becomes singular. However, by de l'Hopital's rule:
$$
\lim_{\hat{\theta}\rightarrow 0} \,\hat{F}'(\hat{\theta}) 
\;=\;
\lim_{\hat{\theta}\rightarrow 0}\,
\frac{ C_1\,+\,
\int_{C_2}^{\hat{\theta}}\hat{\rho}(\hat s)\,\sin^{-2}(\hat s)\,\exp\bigl(k\,\sin^{-2}(\hat{s}) \bigr)\,d\hat s}{
\exp\bigl(k\,\sin^{-2}(\hat{\theta}) \bigr)\,\sin^2(\hat{\theta})}
\;=\;
\lim_{\hat{\theta}\rightarrow 0}\,
\frac{\hat{\rho}(\hat \theta)\,\sin^{-2}(\hat \theta)}{
(1-\cos(2\hat \theta)-2k)\cot(\hat \theta)}
\;,
$$
which vanishes due to the special form of $\hat{\rho}$. Next calculating $\hat{F}''$ from the differential equation $\hat{\mathcal{L}}\hat{F}'=\hat f$,
\begin{align*}
\lim_{\hat{\theta}\rightarrow 0} \,\hat{F}''(\hat{\theta}) 
& =\;
\lim_{\hat{\theta}\rightarrow 0} \,
\frac{\hat{\rho}(\hat{\theta})-\big(\sin^2(\hat{\theta})-2\,k\big)\sin(2\hat{\theta})\,\hat{F}'(\hat{\theta})}{\sin^4(\hat{\theta})}
\;
\;.
\end{align*}
Since both denominator and numerator tend to $0$ for $\hat{\theta}\rightarrow 0$, de l'Hospital's rule can be applied. However, using it directly is not helpful and it is better to first divide both by $\big(\sin^2(\hat{\theta})-2\,k\big)\sin(2\hat{\theta})\hat{G}_0(\hat{\theta})$.
Define $\tilde{\rho}(\hat{\theta})\;=\;\frac{\hat{\rho}(\hat{\theta})}{(\sin^2(\hat{\theta})-2\,k)\sin(2\hat{\theta})}$ and $p(\hat{\theta})=\frac{\sin^4(\hat{\theta})}{(\sin^2(\hat{\theta})-2\,k)\sin(2\hat{\theta})}$ and use that
$(\frac{1}{\hat{G}_0(\hat{\theta})})'\,=\,\frac{1}{\hat{G}_0(\hat{\theta})}\frac{1}{p(\hat{\theta})}$.
Thus one obtains
$$
\lim_{\hat{\theta}\rightarrow 0} \,\hat{F}''(\hat{\theta}) 
\; =\;
\lim_{\hat{\theta}\rightarrow 0} \,\frac{\tilde{\rho}'(\hat{\theta})}{1+p'(\hat{\theta})}\;=\;0\,.
$$
One concludes that the $C^1$-norm of $\hat{F}'$ is uniformly bounded in $\lambda$. Let us note that \cite[Proposition~3]{SS1} presents an iterative procedure showing that also further derivatives of $\hat{F}'$ exist and are bounded.

\vspace{.2cm}

Now let us consider the transformed $\rho(\theta)=\hat{f}\circ t_\lambda^{-1}(\theta)=\exp\bigl(- \,k\,\lambda^{-\frac{1}{2}}\,\cot^2(\theta) \bigr)$. Then $I_{\lambda,N}(\rho)=\hat{I}_N(\hat{\rho})=\Oo(\lambda^{\frac{1}{2}}\|\hat{F}\|_{C^2} )$. Next set $A'=\{\theta\in\SM^1_\pi\,|\,|\theta-\frac{\pi}{2}|\leq\lambda^{\frac{1}{4}}\}$. Then there is a constant $C$ such that $\chi_{A'}\leq C\rho$ and therefore $\nu_\lambda(A')\leq C' \lambda^{\frac{1}{2}}$. On the other hand, it was already shown above that the set $A=\{\theta\in\SM^1_\pi\,|\,\sin^2(2\theta)>\lambda^{\frac{1}{2}}\}$ satisfies $\nu_\lambda(A)=\Oo(\lambda^{\frac{1}{2}})$. Combining these estimates concludes the proof. 
\hfill $\Box$

\begin{coro}
\label{coro-two-fp}
Suppose $\EM[p_2^2]>0$. For every function $f\in C^1(\SM^1_\pi)$ 
$$
\nu_\lambda(f)
\;=\;
f(0)
\;+\;\Oo(\|f\|_{C^1}\,\lambda^{\frac{1}{4}})
\;,
$$
while for $f\in C^2(\SM^1_\pi)$ 
$$
\nu_\lambda(f)
\;=\;
f(0)
\;+\;\Oo(\|f\|_{C^2}\,\lambda^{\frac{1}{2}})
\;.
$$
In particular, every sequence $(\nu_{\lambda_k})_{k\geq 1}$ of invariant measures with $\lambda_k\to 0$ converges weakly to a Dirac peak at the stable fixed point $0$.
\end{coro}

\noindent {\bf Proof:} Let $B=\{\theta\in\SM^1_\pi\,|\,|\theta|>\lambda^{\frac{1}{4}}\}$ and $B^c=\SM^1_\pi\setminus B$. Then by Proposition~\ref{prop-two-fp}
$$
\nu_\lambda(f)
\;=\;
\Oo(\|f\|_\infty\,\lambda^{\frac{1}{2}})
\;+\;
\int_Bf(\theta)\,\nu_\lambda(d\theta)
\;.
$$
Expanding $f(\theta)$ in a Taylor series of order $1$ in $\lambda$ implies the first claim. For the second, let us replace $f$ by $\tilde{f}(\theta)=f(\theta)-\frac{1}{4}f'(0)\sin(4\theta)$ so that $\tilde{f}'(0)=0$. As $\nu_\lambda(\sin(4.))=\Oo(\lambda)$ by the argument preceeding \eqref{eq-dyndirectexp}, one has $\nu_\lambda(\tilde{f})=\nu_\lambda(f)+\Oo(\lambda)$. Now a Taylor series of $\tilde{f}$ to second order in $\lambda$ leads to the second claim.
\hfill $\Box$

\subsection{Evaluation of the Lyapunov exponent and its variance}
\label{sec-hyplyap}

Corollary~\ref{coro-two-fp} can be applied to the function $f(\theta)=\cos(2\theta)$ and implies $I_\lambda(\cos(2\cdot))=1+\Oo(\lambda^\frac{1}{2})$. Furthermore, it was already shown using \eqref{eq-dyndirectexp} that $I_\lambda(\sin(2\cdot))=\Oo(\lambda)$. This allows us to calculate the Lyapunov exponent based on the expansion \eqref{gamma_n} up to order $\lambda$:
$$ 
\gamma_\lambda\;=\;\lambda\,\eta\;+\;\Oo(\lambda^\frac{3}{2}) \,.
$$
This shows the first formula in Theorem~\ref{theo-main}(ii) under the condition $\EM[p_2^2]>0$ on the variance of the entries of $P_n$. The second claim in Theorem~\ref{theo-main}(ii) is an estimate on the variance $\sigma_\lambda$. This will be based on \eqref{eq-sigmalambda}. As it is already known that $g_1$, $g$ and $\gamma_\lambda$ are of order $\Oo(\lambda)$, it is thus sufficient to show that $J_\lambda(g)=\Oo(\lambda^{\frac{1}{2}})$, which results from the following

\begin{proposi}
Assuming $\theta_0=0$. Then for any $f\in C^{3+\alpha}(\SM)$ with $\alpha>0$ 
$$
 J_\lambda(f)\;=\;
\Oo(\lambda^{-\frac{1}{2}})\,.
$$
\end{proposi}

\noindent {\bf Proof.}
Again let us use the rescaled dynamics defined in \eqref{eq-Sexpan}. Let $\hat{\Ll}$ be defined by \eqref{eq-Lhatdef} and set $\hat{f}_\lambda=f\circ t_\lambda$ as above. Furthermore,  let $\hat{\nu}_\lambda$ be the invariant measure of the rescaled dynamics. Then $\nu_\lambda(f)=\hat{\nu}_\lambda(\hat{f})$ and there is a connection between the correlation sums of the two processes:
$$
J_\lambda(f)
\;=\;
\EE_1\Bigl[\sum_{n=1}^\infty\Bigl(f(\theta_n) - {\nu_\lambda}(f)\Bigr)\Bigr]
\;=\;
\EE_1\Bigl[\sum_{n=1}^\infty\Bigl(\hat{f}(\hat{\theta}_n) - {\hat{\nu}_\lambda}(\hat{f}_\lambda) \Bigr)\Bigr]
\;=\;
\hat{J}_\lambda(\hat{f}_\lambda)
\,.
$$
For $\hat{f}\in C^{1+\alpha}(\SM^1_\pi)$ with $\hat{f}=\hat{\mathcal{L}}\hat{F}'$, a Taylor expansion with some  H\"older continuous $\hat{r}$ gives
$$
\hat{J}_\lambda(\hat{F})
\;=\;\hat{F}(\hat{\theta}_0) \,-\,{\hat{\nu}}_\lambda (\hat{F})\,+\, \hat{J}_\lambda(\hat{F})\, -\, \lambda \hat{J}_\lambda(\hat{\mathcal{L}}\hat{F}') \,+\, \lambda^{\frac{3}{2}}\hat{J}_\lambda(\hat{r})\,,
$$
so that
$$
\hat{J}_\lambda(\hat{\mathcal{L}}\hat{F}') 
\;=\;
\frac{1}{\lambda}\big(\hat{F}(\hat{\theta}_0) -{\hat{\nu}}_\lambda(\hat{F})\big)
\,+\,
\lambda^{\frac{1}{2}}\hat{J}_\lambda(\hat{r})\,.
$$
Because $\gamma_\lambda=\Oo(\lambda)$, the {\it a priori} estimate gives $\hat{J}_\lambda(\hat{r})=\Oo(\lambda^{-2})$ and one hence obtains $\hat{J}_\lambda(\hat{\mathcal{L}}\hat{F}')=\Oo(\lambda^{-\frac{3}{2}})$. As mentioned before, every H\"older continuous function with zero at $0$ lies in the range of $\hat{\mathcal{L}}$. Furthermore, $\hat{J}_\lambda$ is invariant under the translation by a constant.  Thus, $\hat{J}_\lambda(\hat{r})=\hat{J}_\lambda(\hat{r}-\hat{r}(0))$.  By \eqref{eq-hyp} the invariant measure $\hat{\nu}_\lambda$ is concentrated at $0$, so the difference $\hat{F}(\hat{\theta}_0) -{\hat{\nu}_\lambda}(\hat{F})$ is of order $\lambda^{\frac{1}{2}}$ because $\theta_0=0$.  Recursively it follows that $\hat{J}_\lambda(\hat{f})\,=\,\Oo(\lambda^{-\frac{1}{2}})$ for every function $\hat{f}\in C^{3+\alpha}(\SM^1_\pi)$. Hence for a function $f_\lambda=\hat{f}\circ t_\lambda^{-1}$ one has
\begin{equation}
\label{eq-Jrel}
J_\lambda (f_\lambda)\;=\;\hat{J}_\lambda(\hat{f})\,=\,\Oo(\lambda^{-\frac{1}{2}})
\;.
\end{equation}
Next for a given smooth function $f$ let us define 
$$
\widetilde{f}(\theta)\;=\;
f(\theta)-f(0)-(f(\tfrac{\pi}{2})-f(0))\sin(t_\lambda^{-1}(\theta))-c_\lambda\,\sin(2\,t_\lambda^{-1}(\theta))\,,
$$
with the constant  defined by 
$$
c_\lambda\;=\;\frac{1}{\pi}\int_{\SM^1_\pi}\frac{f(\theta)-f(0)-(f(\tfrac{\pi}{2})-f(0))\sin(t_\lambda^{-1}(\theta))}{\sin(2\,\theta)}\,d\theta
\;<\;\infty\,.
$$
This constant is uniformly bounded in $\lambda$ as can be seen by analyzing the contributions around the singularities. For example, for small $a>0$ one has $\int^a_{-a}\frac{f(\theta)-f(0)}{\sin(2\,\theta)}d\theta\leq C$ and using that also
$$
\int^a_{-a}\frac{\sin(t_\lambda^{-1}(\theta))}{\sin(2\,\theta)}\,d\theta
\;=\;
\int^a_{-a}\left(\frac{\lambda\tan^2(\theta)}{1+\lambda\tan^2(\theta)}\right)^{\frac{1}{2}}
\frac{1}{\sin(2\,\theta)}\,d\theta
\;\leq\;
\int^a_{-a}\left(\frac{\tan^2(\theta)}{1+\tan^2(\theta)}\right)^{\frac{1}{2}}
\frac{1}{\sin(2\,\theta)}\,d\theta
\;,
$$
is bounded, one sees that also the second summand has a uniformly bounded contribution. Furthermore $\widetilde{f}(0)=\widetilde{f}(\frac{\pi}{2})=0$ and
$$
\int_{\SM^1_\pi}\frac{\sin(2\,t_\lambda^{-1}(\theta))}{\sin(2\theta)}\,d\theta
\;=\;
\int_{\SM^1_\pi}\frac{\lambda^{\frac{1}{2}}}{\lambda\sin^2(\theta)+\cos^2(\theta)}\,d\theta
\;=\;
\int_{\SM^1_\pi}(t_\lambda^{-1})'(\theta)\,d\theta
\;=\;
\pi\,,
$$
so that also $\int_{\SM^1_\pi}\frac{\widetilde{f}(\theta)}{\sin(2\theta)}\,d\theta=0$.  By \eqref{eq-Ldef} one concludes that $\widetilde{f}$ lies in the range of $\Mm$. By \eqref{eq-Jrel} one now concludes $J_{\lambda}(\widetilde{f})=J_{\lambda}(f)+\Oo(\lambda^{-\frac{1}{2}})$ and it thus only remains to analyze the correlation sum of $\widetilde{f}=\Mm F$. Expanding $F\in C^{2+\alpha}(\SM^1_\pi)$ in Taylor series and using the operator $\Mm$ defined in \eqref{eq-Ldef}, one gets
$$
J_\lambda(F)
\;=\;
F(\theta_0) \,-\,{\nu}_\lambda(F)
\,+ \,J_\lambda(F) \,- \,\lambda \,\eta \,J_\lambda(\mathcal{M}F)\, +\, \lambda^2\,J_\lambda(r)\,,
$$
for some H\"older continuous function $r$, since $F\in C^{2+\alpha}(\SM)$. By Proposition~\ref{a-priori} and $\gamma_\lambda=\Oo(\lambda)$, $r$ satisfies the {\it a priori} bound $J_\lambda(r)=\Oo(\lambda^{-2})$. Hence 
$$
J_\lambda(\mathcal{M}F)
\;=\;
\frac{1}{\lambda\, \eta }\big(F(\theta_0) \,-\,{\nu}_\lambda(F) \,+ \,\Oo(1)\big)
\,,
$$
so that  $J_\lambda(\mathcal{M}F)=\Oo(\lambda^{-1})$. By the above applied to $r$ one has $J_{\lambda}(r)=J_{\lambda}(\widetilde{r})+\Oo(\lambda^{-\frac{1}{2}})=\Oo(\lambda^{-1})$. Thus actually  $J_\lambda(\mathcal{M}F)=\Oo(\lambda^{-\frac{1}{2}})$.  This finishes the proof.
\hfill $\Box$


\section{Second order anomalies}
\label{sec-2order}

At a second order anomaly, the term of order $\Oo(\lambda)$ in \eqref{eq-dynexpan} is centered. It is then not necessary to carry out a basis change $M$, but we nevertheless set $P_n=\Pp_n$ and $Q_n(\lambda)=\Qq_n(\lambda)$. Let us begin by recalling some results from \cite{SB,SS1} which are needed below. First of all, one is naturally led to study the second order differential operator on $C^2(\SM^1_\pi)$ defined by
$$
\mathcal{L}
\;=\;
\tfrac{1}{2}\,\EE_n[p_n^2]\partial^2_\theta
\;+\;
\EE_n[q_n+ \tfrac{1}{2}\,p_n p'_n]\partial_\theta
\;,
$$
because for $F\in C^2(\SM^1_\pi)$ one then has
$$
\EE_n\,F(S_{n,\lambda}(\theta))
\;=\; 
F(\theta)\,+\,\lambda^2\,(\Ll F)(\theta)\,+\,\Oo(\lambda^3)
\;.
$$
From this one can again control Birkhoff sums in the range of $\Ll$. For sake of simplicity (weaker hypothesis are possible), it will be assumed that $\EE_n[p_n^2]>0$. This assures that $\Ll$ is elliptic. Let also introduce the formal adjoint $\mathcal{L}^*=\partial^2_\theta \tfrac{1}{2}\,\EE_n[p_n^2]-\partial_\theta \EE_n[q_n+ \tfrac{1}{2}\,p_n p'_n]$. Then \cite[Theorem~3]{SS1} implies the following result.  Here scalar product and orthogonal complement are taken in $L^2(\SM^1_\pi)$.

\begin{theo}
\label{theo-oldres}
Under the above ellipticity hypothesis, one has  $\mbox{\rm dim}(\mbox{\rm Ran}({\mathcal{L}})^\perp)=1$ and there is a unique smooth and non-negative function $\rho\in \mbox{\rm Ker}(\mathcal{L}^*)$ with unit integral. For every function $f\in C^1(\SM^1_\pi)$ one has 
$$
I_{\lambda,N}(f)\;=\;\langle \rho|f\rangle\,+\,\Oo(\lambda,\,(\lambda N)^{-1}) \,,
$$
where $\langle \rho|f\rangle=\int_{\SM^1_\pi}  f(\theta)\,\rho(\theta)\, d\theta$.
\end{theo}

Using Theorem~\ref{theo-oldres} and \eqref{eq-gamma1} as well as \eqref{gamma_n}, it is straightforward to  write out the expansion for the Lyapunov exponent given in Theorem~\ref{theo-main}(iii) with rigorous control on the error terms. The formula for $C_s$ was already given in \cite{SB} and invokes $\rho$ which, in general, cannot be caluclated explicitly. For the calculation of the variance, one needs again to control $J_\lambda(f)$.

\begin{lemma}
\label{lem-secano}
For a function $f\in C^{5+\alpha}(\SM_\pi^3)$ with $\alpha>0$ the correlation sum $J_\lambda(f)$ is given by
$$
J_\lambda(f)\;=\;-\frac{1}{\lambda^2}(F(\theta_0)-\nu_\lambda(F)\,+\,\Oo(\lambda^{-1}) 
\;,
$$
with a periodic function $F=\mathcal{L}^{-1}(f - \langle f|\rho \rangle)$ defined uniquely up to a constant. 
\end{lemma}

\noindent {\bf Proof.} Let us first recall \cite{SS1} that the range $\mbox{\rm Ran}({\mathcal{L}})$ includes all smooth functions that are orthogonal to $\rho$ in the $L^2$-sense. Hence $F$ is well-defined. Furthermore, $J_\lambda$ is invariant under shifts by a constant. Therefore  $J_\lambda(f)=J_\lambda(\Ll F)$. As in the proof of Lemma~\ref{lem-Jell} a Taylor expansion of $F\in C^{3+\alpha}(\SM^1_\pi)$ shows
$$
J_\lambda(F)
\; = \; 
F(\theta_0) - \nu_\lambda(F)\,+\,J_\lambda(F)\,+\,\lambda^2 J_\lambda(\mathcal{L}F)\,+\,\lambda^3 J_\lambda(r)\,,
$$
where $r$ is a H\"older-continuous residual function for which an priori bound $J_\lambda(r)=\Oo(\lambda^{-3})$ holds. From this one deduces $J_\lambda(\mathcal{L}F)=\Oo(\lambda^{-2})$ and hence $J_\lambda(f)=\Oo(\lambda^{-2})$. As all this equally well applies to $r$, namely $J_\lambda(r)=\Oo(\lambda^{-2})$, the equation actually already implies the claim.
\hfill $\Box$

\vspace{.2cm}

The calculation of $\sigma_\lambda$ will again be based on \eqref{eq-sigmalambda}. By \eqref{gamma_n} one has $g_n=\lambda h_n+\lambda^2 f_n+\Oo(\lambda^3)$ with certain trigonometric polynomials $h_n$ and $f_n$. By hypothesis, one has $\EM[h_n]=0$. Hence the function $g=\EE_n[g_n]$ has an expansion given by $g=\lambda^2 f+\Oo(\lambda^3)$ with $f=\EM[f_n]$. Let $F=\Ll^{-1}(f-\langle \rho|f\rangle)$. Cancelling out the $\lambda$-factors, one then concludes from Lemma~\ref{lem-secano}
$$
J_\lambda(g)\;=\;\nu_\lambda(F)-F(\theta_0)\,+\,\Oo(\lambda)\,. 
$$
As $\gamma_\lambda=\Oo(\lambda^2)$, one thus has 
$$
\sigma_\lambda
\;=\;
\EE_{\nu_\lambda}
\EM_1 \left[\lambda^2 h_1^2
\,+\,(\lambda\,h_1+\lambda^2 f_1)\, J_\lambda(g)
\right]\,+\,\Oo(\lambda^3)
\;.
$$
Replacing $J_\lambda(g)$ one sees that due to $\EM[h_1]=0$ the variance satisfies $\sigma_\lambda=\Oo(\lambda^2)$. In principle the above formula gives all the lowest order contributions, provided the contribution of order $\Oo(\lambda) $ to $J_\lambda(g)$ can be calculated. This provides $C'_s$ and it appears to be a coincidence for it to be equal to $C_s$.

\section{Examples}
\label{sec-examples}

\subsection{Harmonic chain}

A harmonic chain consists of sequence of masses $m_n>0$ with displacements $u_n$ from it's equlibrium position for which the stationary equation of motion at frequency $\omega$  is
\begin{equation*}
 -m_n\omega^2 u_n\,=\,u_{n+1}-2 u_n + u_{n-1}\,,
\qquad
n\in\ZM
\;.
\end{equation*}
The masses are supposed to be i.i.d. randomly distributed with a distribution of compact support bounded away from zero. This is a Jacobi matrix and the associated  transfer matrix is given by
\begin{equation*}
 \mathcal{T}_{n,\omega}\;=\;\begin{pmatrix} 2-\omega^2 m_n & -1 \\ 1 & 0 \end{pmatrix}\,.
\end{equation*}
The transfer matrix is in $\mbox{Sl}(2,\RM)$ and it is conjugate to a rotation. The aim will now be to study the Lyapunov exponent and variances of the associated random products perturbatively in $\omega$. For small $\omega$, $\mathcal{T}_{n,\omega}$ lies in the vicinity of the Jordan block. Indeed, by the basis change with $M_1=\begin{pmatrix}1 & 0 \\1 & -1\end{pmatrix}$ one obtains
\begin{equation*}
 M_1 \mathcal{T}_{n,\omega} M_1^{-1}\;=\;
\begin{pmatrix}
1 & 1 \\
0 & 1 
\end{pmatrix}
-m_n\omega^2
\begin{pmatrix}
1 & 0\\
1 & 0
\end{pmatrix}\,.
\end{equation*}
Now one blows up the vicinity of the stable point by conjugation with the matrix $M_2=\begin{pmatrix}\omega & 0 \\0 & 1\end{pmatrix}$
\begin{equation*}
M_2 M_1 \mathcal{T}_{n,\omega} (M_2 M_1)^{-1}\;=\;
\one\,+\,\omega
\begin{pmatrix}
0 & 1\\
-m_n & 0
\end{pmatrix}
\,-\,\omega^2
\begin{pmatrix}
m_n & 0\\
0 & 0
\end{pmatrix}.
\end{equation*}
The last basis change is done with the matrix 
$M_3\,=\,\begin{pmatrix}-\sqrt{\EE[m_n]} & 0 \\ 0 & 1\end{pmatrix}$
\begin{equation*}
T_{n, \omega}\;=\;M_3 M_2 M_1 \mathcal{T}_{n,\omega} (M_3 M_2 M_1)^{-1}
\;=\;
\one\,+\,\omega
\begin{pmatrix}
0 & -\sqrt{\EE[m_n]}\\
\frac{m_n}{\sqrt{\EE[m_n]}} & 0
\end{pmatrix}
\,-\,\omega^2
\begin{pmatrix}
m_n & 0\\
0 & 0
\end{pmatrix}\,.
\end{equation*}
This is a first order elliptic order model already in the form of \eqref{elliptic}. Therefore the results of Section~\ref{sec-ellfirst} giving the constant in Theorem~\ref{theo-main}(i) shows
\begin{equation*}
\gamma_\omega\;=\;\frac{\omega^2}{8}\,\frac{\EE[m_n^2]-\EE[m_n]^2}{\EE[m_n]}\,+\,\Oo(\omega^3)\;=\;\sigma_\omega\,+\,\Oo(\omega^3)\,.
\end{equation*}
The formula for the Lyapunov exponent is known at least since the work of \cite{Ish}.

\subsection{Band edges of the Anderson model}

The one-dimensional discrete Anderson model is described by the stationary Schr\"odinger equation 
\begin{equation}
 E\psi_n\;=\;\psi_{n+1} + \psi_{n-1}\,+\,\lambda v_n \psi_n\,.
\end{equation}
Here $\psi_n$ are the probability amplitude, $E\in\RM$ is the energy and the $v_n\in\RM$ are i.i.d. potential energies according to a centered distribution of compact support. The transfer matrix is given by
\begin{equation*}
 \mathcal{T}_{n,\omega}\;=\;\begin{pmatrix} E-\lambda v_n & -1 \\ 1 & 0 \end{pmatrix}\,.
\end{equation*}
The case $E=0$ corresponds immediately to an anomaly once one takes two consecutive transfer matrices as the basic building blocks \cite{KW,SB}. Here we will rather focus on the band edges $|E|=2$, and show that this also leads to an anomaly \cite{DG,SS1}.  Let us focus on a situation where $E=2+w\lambda$ for some $w\in\RM$, again with the aim of calculating the Lyapunov exponent and variance perturbatively in $\lambda$. Their behavior depends strongly on the sign of $w$. If $w<0$, one has an elliptic first order anomaly. From the technical point of view, this problem can be dealt with the same procedures as the previous one substituting $\omega^2$ by $\lambda$ and $\EE[m_n]$ by $-w$, and this leads to
\begin{equation*}
\gamma_\lambda\;=\;\frac{\lambda}{8}\,\frac{\EE[v_n^2]-w^2}{w}\;+\;\Oo(\lambda^{\frac{3}{2}})\;=\;\sigma_\lambda\;+\;\Oo(\lambda^{\frac{3}{2}})\,.  
\end{equation*}
For $w>0$ one has a hyperbolic first order anomaly. The basis changes into the normal form \eqref{eq-hypcoord} are consecutively given by the matrices  $M_1=\begin{pmatrix}1 & 0 \\1 & -1\end{pmatrix}$,  $M_2=\begin{pmatrix}\lambda^{\frac{1}{2}} & 0 \\0 & 1\end{pmatrix}$
and $M_3=\begin{pmatrix} w^{\frac{1}{2}} & 1 \\ -w^{\frac{1}{2}} & 1\end{pmatrix}$ respectively. Now the results of Section~\ref{sec-hyperbolic} show
\begin{equation*}
 \gamma_\lambda\;=\;\lambda^{\frac{1}{2}}w^{\frac{1}{2}}\,+\,\Oo(\lambda^{\frac{3}{4}})\,,
\qquad
\sigma_\lambda\;=\;\Oo(\lambda^{\frac{3}{4}})
\;.
\end{equation*}
%

\subsection{Random Kronig-Penney model}

The random Kronig-Penny model describes a quantum particle by a continuous one-dimensional Schr\"odinger operator with singular random potentials on $\ZM$. This model also reduces to the study of products of random transfer matrices, as described in detail in \cite{DKS}. Let us just recall from \cite{DKS} the special from of the transfer matrices close to the so-called critical energies  $E_l\,=\,(\pi l)^2$, $l\in\mathbb{N}$, of the system, namely for $\epsilon\geq 0$ and after adequate basis changes
\begin{equation*}
T^{E_l -\varepsilon}_n
\;=\;
R_{-\eta\,\varepsilon^{\frac{1}{2}}}
\left[\one\,+\,
\varepsilon^{\frac{1}{2}}\, \frac{\widetilde{v}_n}{2\sqrt{2\bar{v} E_l}}
\begin{pmatrix} -1 & 1 \\ -1 & 1 \end{pmatrix}
-
\varepsilon\,\frac{\bar{v}}{4 E_l} \,
\begin{pmatrix} 0 & 1 \\ 1 & 0 \end{pmatrix}
-
\frac{\varepsilon\,\widetilde{v}_n}{2E_l}
\begin{pmatrix} 0 & 1 \\ 1 & 0 \end{pmatrix}
\right]
+\mathcal{O}(\varepsilon^{\frac{3}{2}})
\,,
\end{equation*}
and 
\begin{equation*}
T^{E_l +\varepsilon}_n
\;=\;
\begin{pmatrix} 1-\eta\varepsilon^{\frac{1}{2}} & 0 \\ 0 & 1+\eta\varepsilon^{\frac{1}{2}} \end{pmatrix}
\,+\,
\varepsilon^{\frac{1}{2}}\, \frac{\widetilde{v}_n}{2\sqrt{2\bar{v} E_l}}
\begin{pmatrix} -1 & 1 \\ -1 & 1 \end{pmatrix}
\,+\,
\varepsilon\,\frac{v_n}{4 E_l} \,
\begin{pmatrix} 1 & 1 \\ 1 & 1 \end{pmatrix}
\,+\,\mathcal{O}(\varepsilon^{\frac{3}{2}})
\;.
\end{equation*}
Here the $v_n\in\RM$ are again i.i.d. with a distribution that is assumed to have compact support (not concentrated in a single point) with expectation value $\bar{v}$ and deviations $\widetilde{v}_n=v_n-\bar{v}$.  Finally $R_{-\eta\,\varepsilon^{\frac{1}{2}}}$ is a rotation with an angle $\eta\;=\;\sqrt{\frac{\bar{v}}{2E_l}}$. Now $T^{E_l -\varepsilon}_n$ is again an elliptic first order anomaly with $\varepsilon^{\frac{1}{2}}$ playing the role of $\lambda$. Then the Lyapunov exponent $\gamma^{E_l-\varepsilon}$ and its variance $\sigma^{E_l-\varepsilon}$ are up to the order $\varepsilon^{\frac{3}{2}}$ given by 
\begin{equation*}
\gamma^{E_l-\varepsilon}
\;=\;
\frac{\EE(v_n^2-\bar{v}^2)}{16\,\bar{v} E_l}\,\varepsilon\;+\;\Oo(\varepsilon^{\frac{3}{2}})
\;=\;
\sigma^{E_l-\varepsilon}\,+\,\Oo(\varepsilon^{\frac{3}{2}})\,.
\end{equation*}
On the other hand, $T^{E_l +\varepsilon}_n$ is a hyperbolic first order anomaly and thus the above results imply
\begin{equation*}
\gamma^{E_l+\varepsilon}
\;=\;
\left(\frac{\bar{v}}{2E_l}\right)^{\frac{1}{2}}\,\varepsilon^{\frac{1}{2}}\;+\;\Oo(\varepsilon^{\frac{3}{4}})\,,
\qquad
\sigma^{E_l+\varepsilon}\;=\;\Oo(\varepsilon^{\frac{3}{4}})
\;.
\end{equation*}
The Lyapunov exponents were already obtained by the authors in the previous work \cite{DKS}, but the proofs and the error estimates there are given too briefly.  


\end{document}